\documentclass[12pt,reqno]{article}

\usepackage{amsmath, amsthm, amssymb,epsfig,alltt}

\def\a{\alpha}
\def\b{\beta}
\def\d{{\delta}}
\def\l{\lambda}
\def\e{\epsilon}
\def\p{\partial}
\def\k{{\kappa}}
\def\m{\mu}
\def\n{\nu}
\def\t{\tau}
\def\th{\theta}
\def\s{\sigma}
\def\g{\gamma}

\def\half{\frac{1}{2}}

\def\whatphi{{\widehat{\phi}}}

\def\barpsi{{\bar \psi}}

\def\nn{\nonumber}

\def\2pap{2\pi\alpha^\prime}

\def\beq{\begin{eqnarray}}
 \def\eeq{\end{eqnarray}}
 \def\4pap{4\pi\a^\prime}

 \def\bolG{{\boldsymbol G}}
 \def\bbZ{\mathbb{Z}}

\begin{document}


\title{\bf Hetero-Junction of Two Quantum Wires: \\ 
Critical Line and Duality}

\author{Taejin Lee \\~~\\
Department of Physics, Kangwon National University, \\
Chuncheon 200-701 Korea }

\maketitle

\centerline{\bf Astract}
Applying the Fermi-Bose equivalence and the boundary state formulation, we study the hetero-junction of 
two quantum wires. Two quantum wires are described by Tomonaga-Luttinger (TL) liquids with different
TL parameters and electrons transport between two wires is depicted by a simple hopping interaction. 
We calculate the radiative corrections to the hopping interaction and obtain the renormalization (RG) 
exponent, making use of the perturbation theory based on the boundary state formulation. 
The model exhibits a phase transition at zero temperature. We discuss the critical line which 
defines the phase boundary on the two dimensional parameter space. The model also exhibits
the particle-kink duality, which maps the strong coupling regime of the model
onto the weak coupling regime of the dual model. The strong coupling regime of the model 
is found to match exactly the weak coupling regime of the dual model. 
This model is also important to study the critical behaviors of the two dimensional dissipative 
system with anisotropic friction coefficients.




\vskip 0.5cm


\vskip 2cm

\section{Introduction}

The junctions of quantum wires \cite{wong1993, fendley1995, Safi95, chamon2003, oshikawa2006, Hou2008, aristov2011, Hou2012, aristov2013}
are the basic building blocks of quantum circuits \cite{larkin1997, fazio2001, Oskin2003, sodano2005, Sarma2007, Alicea2011}, which may bring the quantum 
computation \cite{shor1996} into reality. When we join two quantum wires, which described by the Tomonaga-Luttinger \cite{Tomonaga1950,Luttinger1963} (TL) liquids,
together to make a circuit, the TL parameters of two wires, $\a_1$ and $\a_2$,  
may be different in general. In the present work we 
study the hetero-junction of two quantum wires in detail, making use of the boundary state formulation 
\cite{callan90, Hassel2006, Leechiral2015, Leemulti2016}.
The hetero junction of two quantum wires has been discussed in ref. \cite{Safi95} and later 
it has been extended to the hetero-junction of three wires in ref. \cite{Hou2012}. 
However, the critical behaviors of the model has not been fully explored. Here we will directly calculate the 
radiative corrections to the hopping interaction and discuss the particle-kink duality (the Schmid duality) 
\cite{schmid1983, guinea1985, fisher1985}
in the context of the junction of two quantum wires, which maps the strong coupling regime of the particle 
theory onto the weak coupling regime of its dual theory. 
Depending on the effective TL parameter, which is given by $\a_{\rm eff} = \frac{2}{\a^{-1}_1+ \a^{-1}_2}$, 
the model exhibits a phase transition at zero temperature. This result is in perfect agreement with the 
previous works \cite{Safi95, Hou2008}.
 
The model of the hetero-junction is similar to the spin-dependent TL model with a single barrier discussed in refs.\cite{furusaki1993a,LeeU(1)2015}.
The spin-dependent TL model is described by two TL liquids; one for the charge degree of 
freedom and the other for 
the spin degree of freedom. Since due to the spin-charge separation in one dimension, the TL parameters of two TL liquids for the spin and charge degrees of freedom may differ from each other. The hopping interaction between two wires 
may play the role of the single barrier. Thus, we expect that the two models may share common critical behaviors. On the other hand, 
the model of hetero-junction is still different from the spin-dependent TL model in that the electrons
transfer through only one channel while the spin-dependent TL model has two channels of transport. 

This model is also important to study the critical behaviors of the two dimensional dissipative system of the 
Caledeira-Legget type \cite{caldeira83ann} 
with anisotropic friction coefficients. Integrating out the 
bulk boson fields may result in the non-local dissipative action of the Caldeira-Legget type on the 
boundary. The two different TL parameters transcribe into two different friction coefficients of the 
anisotropic dissipative system. 

In order to calculate the radiative corrections and the RG exponent, we adopt the 
boundary state formulation \cite{callan90, Hassel2006, Leechiral2015, Leemulti2016, LeeU(1)2015},
which easily incorporates with the relativistic quantum field theory in (1+1) dimensions \cite{brown1992}.  
We bosonize the model by using the Fermi-Bose equivalence and refermionize the model as a Thirring 
model with a boundary interaction to define a perturbation theory. 
Making use of the vector and axial $U(1)$ local phase transformations, 
we encode all the non-trivial interactions 
on the boundary. As a result the $U(1)$ axial phase boson fields become dynamical and 
the bulk action contain their free field action. 
The radiative corrections can be evaluated by usual Feynman diagram expansion. 
The free bulk action defines the free propagators of the fermion fields and boson fields on the boundary. 
In a recent work \cite{LeeU(1)2015} 
we have shown that the boundary state formulation yields the correct RG exponent
of the spin-dependent TL model. 

In the strong coupling regime, the particles are mostly localized at the minima of the boundary periodic potential,
which represents the hopping interaction between two wires. The boundary state becomes the Dirichlet state. corresponding to total reflection. 
Electrons no longer transfer between two wires in the localized phase of the 
strong coupling regime. 
However, it is well known that this type of model with a boundary periodic potential 
has dual degrees of freedom in the strong coupling regime. The kinks, which are classical solutions interpolating between local minima of the periodic potential become dynamical. 
Their collective motions may be described by a dual field, of which action takes exactly the 
same form as that of the particle field but with 
weak coupling constants. This is called the particle-kink duality or Schmid duality. In order to have 
a stable kink solution, we may need a kinetic term for the boson field on the boundary. 
The kinetic term of the boson theory, which represented by four fermi interaction terms in the fermion
theory, is absent at the tree level action. But the perturbation theory implies that such term 
may arise as a finite one-loop correction to the four fermi interaction. 
We discuss the particle-kink duality in the context of
the hetero-junction of two quantum wires. Since the model has only one channel of charge transport, 
there is only one kind of kink. As a consequence of it, the dual model contains one component dual field only. 
In the dual theory the hetero-junction is described by a single TL liquid with an effective TL parameter 
$\a_{\rm dual} = 1/{\a_{\rm eff}}= \frac{\a^{-1}_1+ \a^{-1}_2}{2}$.

\section{Model of Hetero-Junction of Two Quantum Wires}

The hetero-junction of two quantum wires may be described by the following Euclidean action at finite temperature 
\beq
S &=& \frac{1}{2\pi} \int_0^\infty dx \int^{\b_T/2}_{-\b_T/2} dt \,\, \sum_{a=1}^2\Biggl\{
\barpsi^a \left(\bar\g^0 \p_t + v \bar\g^1 \p_x \right) \psi^a + 
\bar g_2^a \left(\psi^{a\dag}_L \psi_L^a \psi^{a\dag}_R \psi_R^a \right) \nn\\
&&
+ \frac{\bar g_4^a}{2} \left[ \left(\psi^{a\dag}_L \psi_L^a \right)^2 + \left(\psi^{a\dag}_R \psi_R^a \right)^2 \right]
\Biggr\} \nn\\
&& + \frac{\bar V_0}{4\pi} \int^{\b_T/2}_{-\b_T/2} dt \,  \left(\psi^{1\dag}_L \psi^{2}_L 
+\psi^{2\dag}_L \psi^{1}_L -\psi^{1\dag}_R \psi^{2}_R -\psi^{2\dag}_R\psi^{1}_R 
\right)\Bigl\vert_{x=0} ,
\eeq
where $\b_T = 1/T$ and 
\beq
\bar \g^0 = \s_2= \left(\begin{array}{cc}
  0 & -i \\
  i & 0 
\end{array}\right), ~~ \bar \g^1 = \s_1 = \left(\begin{array}{cc}
  0 & 1 \\
  1 & 0 
\end{array}\right).
\eeq
The action contains two four-fermi interaction terms, corresponding to the forward scattering.  
Two quantum wires are depicted by two different TL liquids. Thus, the coupling constants of one wire
$(\bar g^1_2, \bar g^1_4)$ may differ from the coupling constants of the other wire,
$(\bar g^2_2, \bar g^2_4)$ . The last boundary term is the hopping interaction, through which 
the charges transfer from one wire to the other.

In order to apply the Fermi-Bose equivalence, we redefine the model on a cylindrical space-time of which 
boundary is a unit circle \cite{LeeTL2015}. 
The world sheet coordinates, $\t$ and $\s$ are given by
\beq
\t = \frac{2\pi}{v \b_T}\, x, ~~~ \s = \frac{2\pi}{\b_T}\, t .
\eeq
At the same time we also scale the fermion fields as $\psi^a \rightarrow \psi^a/\sqrt{v}$, 
$\psi^{a\dag} \rightarrow \psi^{a\dag}/\sqrt{v}$. 
Making use of the Fermi-Bose equivalence \cite{Lee2009q,Leeklein2015}
\begin{subequations}
\beq
\psi^1_L &=& e^{-\frac{\pi}{2} i \left(p^1_L + p^1_R  \right)} e^{-\sqrt{2} i \varphi^1_L}, \label{n3fermion1}\\
\psi^2_L &=& e^{-\frac{\pi}{2} i \left(p^2_L+ 2 p^1_L + p^2_R + 2p^1_R \right)} e^{-\sqrt{2} i \varphi^2_L}, \label{n3fermion2}\\
\psi^1_R &=& e^{-\frac{\pi}{2} i \left(p^1_L + p^1_R  \right)} e^{\sqrt{2} i \varphi^1_R}. \label{n3fermion3}, \\
\psi^2_R &=& e^{-\frac{\pi}{2} i \left(p^2_L+2 p^1_L+ p^2_R+ 2p^1_R\right)} e^{\sqrt{2} i \varphi^2_R}, \label{n3fermion4}
\eeq
\end{subequations}
and the Neumann condition in the fermion theory 
\beq
\psi^a_L \vert {\bf N} \rangle = i \psi^{a\dag}_R \vert {\bf N} \rangle, ~~~
\psi^{a\dag}_L \vert {\bf N} \rangle = i \psi^{a}_R \vert {\bf N} \rangle,~~~ a = 1, 2,
\eeq
or the Neumann condition in the boson theory
\beq
\varphi^a_L \vert {\bf N} \rangle = \varphi^a_R \vert {\bf N} \rangle, ~~~ a =1, 2, 
\eeq
we may rewrite the hopping interaction between the quantum wires in terms of the boson fields 
$\varphi^1$ and $\varphi^2$ as 
\beq
\frac{V_0}{2\pi}\left( e^{i \frac{\varphi^1-\varphi^{2}}{\sqrt{2}}}+ 
e^{-i \frac{\varphi^1-\varphi^{2}}{\sqrt{2}}} \right). \label{boundaryz}
\eeq
The bulk action can be also bosonized by making use of the Fermi-Bose equivalence 
Eqs.(\ref{n3fermion1},\ref{n3fermion2},\ref{n3fermion3},\ref{n3fermion4}) or 
\begin{subequations}
\beq
:\psi^a_L \psi^a_L: &=& \frac{i}{\sqrt{2}}\left(\p_\t-i\p_\s \right) \varphi^a_L =  
\frac{i}{\sqrt{2}}\left(\p_\t-i\p_\s \right) \varphi^a, \label{fermibosea}\\
:\psi^a_R \psi^a_R: &=& -\frac{i}{\sqrt{2}}\left(\p_\t+i\p_\s \right) \varphi^a_R =  
-\frac{i}{\sqrt{2}}\left(\p_\t+i\p_\s \right) \varphi^a . \label{fermiboseb}
\eeq
\end{subequations}
The bulk action takes the following form in terms of the boson fields
\beq
S_{\rm bulk} &=& \frac{\a_1}{4\pi}\int d\t d\s  \p \varphi^1 \p \varphi^1 + 
\frac{\a_2}{4\pi}\int d\t d\s  \p \varphi^2 \p \varphi^2, \label{bulkz}
\eeq
The TL parameters $\a_1$ and $\a_2$ are related to the coupling constants of the four fermi 
interactions as \cite{LeeTL2015}
\beq
\a_1 = \half \sqrt{(1+ g^1_2)^2 - (g^1_4)^2}, ~~~\a_2 = \half \sqrt{(1+ g^2_2)^2 - (g^2_4)^2}
\eeq
where 
\beq
g^a_2 = \frac{\b^2_T}{4\pi^2 v } \bar g^a_2, ~~~ 
g^a_4 = \frac{\b^2_T}{4\pi^2 v } \bar g^a_4, ~~~ a = 1 ,2 .
\eeq
Collecting the boundary action Eq.(\ref{boundaryz}) and the bulk action Eq.(\ref{bulkz}), 
we have the boson action of the 
hetero-junction as follows
\beq
S &=& \frac{\a_1}{4\pi}\int d\t d\s  \p \varphi^1 \p \varphi^1 + 
\frac{\a_2}{4\pi}\int d\t d\s  \p \varphi^2 \p \varphi^2 \nn\\
&&+ \frac{V_0}{2\pi} \int d\s \left(e^{\frac{i}{\sqrt{2}} (\varphi^1 - \varphi^{2})}
+ e^{-\frac{i}{\sqrt{2}} (\varphi^1 - \varphi^{2})} \right)\Bigl\vert_{\t=0} .
\eeq

It may be convenient to take an $SO(2)$ rotation, before proceeding with the perturbation theory
\beq
\begin{pmatrix} \varphi^1 \\ \varphi^2 \end{pmatrix} = \frac{1}{\sqrt{2}} \begin{pmatrix} 
~~1 & 1 \\ -1 & 1 \end{pmatrix} \begin{pmatrix} \phi^1 \\ \phi^2 \end{pmatrix} .
\eeq
In terms of the boson fields $(\phi^1, \phi^2)$ the action of the model is written as 
\beq
S = \frac{1}{4\pi} \int d\t d\s \p \phi^a \p \phi^a + \frac{1}{4\pi} \int d\t d\s 
\p \phi^a g_{ab} \p \phi^b + 
\frac{V_0}{2\pi} \int d\s \left(e^{i\phi^1}+ e^{-i\phi^1}\right) \label{bosonactionq}
\eeq
where 
\beq
g_{ab} = \begin{pmatrix} \frac{\a_1+\a_2}{2} -1 & \frac{\a_1-\a_2}{2} \\ 
\frac{\a_1-\a_2}{2} & \frac{\a_1+\a_2}{2} -1 \end{pmatrix}.
\eeq
When $\a_1 = \a_2 = 1$, the model reduces to the Scmid model at the critical point, which 
is exactly solvable
\beq
S = \frac{1}{4\pi} \int d\t d\s \p \phi^a \p \phi^a + 
\frac{V_0}{2\pi} \int d\s \left(e^{i\phi^1}+ e^{-i\phi^1}\right). \label{critical}
\eeq
We will treat the second and third terms of the boson action Eq.(\ref{bosonactionq}) 
as interaction terms and develop a perturbation theory.  

In order to understand that the model is critical and exactly solvable at $\a_1=\a_2=1$,
we need to introduce auxiliary boson fields $\bar\phi^a$, $a=1, 2$. Defining four boson fields 
$\Phi^a{}_i$, $a=1, 2$, $i=1,2$ as follows 
\beq
\Phi^a{}_{1} = \frac{1}{\sqrt{2}}\left(\bar\phi_a+ \phi_a\right),~~~
\Phi^a{}_{2} = \frac{1}{\sqrt{2}}\left(\bar\phi_a -\phi_a\right), ~~~ a =1, 2,
\eeq
we may rewrite the boson action Eq.(\ref{bosonactionq}) as 
\beq
S = \frac{1}{4\pi} \int d\t d\s \sum_{a,i =1}^2 \p \Phi^a{}_i \left(\delta_{ab}+g_{ab} 
\right) \p \Phi^b{}_i + 
\frac{V_0}{4\pi} \int d\s \sum_{i=1}^2 \left(e^{i\sqrt{2}\Phi^1{}_i}+ 
e^{-i\sqrt{2}\Phi^1{}_i}\right). \label{bosonactionnew}
\eeq
We choose the Dirichlet boundary condition for $\bar\phi^a$, $a=1 ,2$, so that the boundary terms 
of Eq.(\ref{bosonactionnew}) reduce
to the boundary hopping interaction terms of Eq.(\ref{bosonactionq}).
We note that the boson fields $\bar\phi^a$, $a=1, 2$, 
do not couple to the physical boson fields $\phi^a$, $a=1, 2$ and they can be safely integrated out
\cite{Lee2008}. 
Thus, they are auxiliary. However, they play a crucial role in developing a perturbation theory of the 
model as we shall see. 
        
Introducing the auxiliary boson fields $\bar\phi^a$, $a=1, 2$, enables us to refermionize the model.
Using the Fermi-Bose equivalence, we may write new fermion fields $\psi^a{}_i$, $a=1, 2$, $i = 1, 2$
in terms of the boson fields $\Phi^a{}_i$, $a=1, 2$, $i= 1, 2$ as 
\beq
\psi^a{}_{1L} &=& \eta^a{}_{1L} :e^{-i\sqrt{2} \Phi^a{}_{1L}}:, ~~~
\psi^a{}_{2L} = \eta^a{}_{2L} :e^{-i\sqrt{2} \Phi^a{}_{2L}}: \\
\psi^a{}_{1R} &=& \eta^a{}_{1R} :e^{i\sqrt{2} \Phi^a{}_{1R}}:, ~~~
\psi^a{}_{2R} = \eta^a{}_{2R} :e^{i\sqrt{2} \Phi^a{}_{2R}}:, ~~a = 1, 2, \nn
\eeq
where 
and $\eta^a{}_{iL}$, $\eta^a{}_{iR}$ are Klein factors which ensure the anti-commutation 
relationship between the fermion fields. 
For an explicit construction of the Klein-factors, we may refer to ref.\cite{Leeklein2015}. 
Making use of the Fermi-Bose equivalence Eqs.(\ref{fermibosea}, \ref{fermiboseb}) again,
we may rewrite the  boson action Eq.(\ref{bosonactionnew}) as 
\beq\label{fermionactionw}
S &=& \frac{1}{2\pi} \int d\t d\s \left(\sum_{a,i} \bar\psi^a{}_{i} \g^\m \p_\m \psi^a{}_{i} + 
\sum_{a,b,i}\frac{g^{ab}}{4} j^\m{}_{ai} j_\m{}_{bi} \right) 
+ \frac{V_0}{4\pi} \int d\s \sum_{i=1}^2 \bar\psi^1{}_{i} \psi^1{}_{i} \Bigl\vert_{\t=0} ~ 
\eeq
where $j^\m{}_{ai} = \bar\psi_{ai} \gamma^\m \psi_{ai}$ and 
\beq
\g^0= \bar\g^1 = \s^1, ~~~~ \g^1 = \bar\g^0 = \s^2, ~~~ \g^5 = -i\g^0 \g^1 = \s^3 .
\eeq
This is a Thirring model with four flavors and boundary mass terms. At the critical point where
$\a_1 =\a_2 =1$, the model reduces to a free fermion theory with boundary fermion bilinears
\beq
S &=& \frac{1}{2\pi} \int d\t d\s \sum_{a,i=1}^2 \bar\psi^a{}_{i} \g^\m \p_\m \psi^a{}_{i} 
+ \frac{V_0}{4\pi} \int d\s \sum_{i=1}^2 \bar\psi^1{}_{i} \psi^1{}_{i} \Bigl\vert_{\t=0}. 
\eeq
Thus, it becomes manifest that the model is critical and exactly solvable at the  point, 
$\a_1=\a_2=1$ when it is rewritten as a fermion theory with the help of the auxiliary 
boson fields $\bar\phi^a$, $a=1, 2$ and the Fermi-Bose equivalence. 

\section{Perturbation Analysis of The Model}

The Thirring action Eq.(\ref{fermionactionw}) may be rewritten as follows, if four Abelian $U(1)$ vector fields 
$A_{i\m a}$, $i, a = 1, 2$ are introduced
\beq
S &=& \frac{1}{2\pi} \int d\t d\s \sum_{a,b,i} \left[\bar\psi^a{}_{i} \g^\m \left(\p_\m +iA_{i\mu a}\right)\psi^a{}_{i}+ A_{i\m a} (g^{-1})_{ab} A^\m{}_{i b}\right]  \nn\\
&&+ \frac{V_0}{4\pi} \int d\s \sum_{i=1}^2 \bar\psi^1{}_{i} \psi^1{}_{i}\Bigl\vert_{\t=0}. 
\eeq
Decomposing the Abelian vector fields 
\beq
A^\m{}_{ia} = \e^{\mu\n}\p_\nu \th_{ia}+ \p^\m \chi_{ia}, \quad i=1, 2, \quad a=1, 2,
\eeq
we can remove the interaction between the vector fields and the fermion fields in the bulk action  
by a $(U_V(1))^2 \times (U_A(1))^2$ local phase transformation,
\beq
\psi^a{}_{i} = e^{-i\g_5 \th^a{}_{i} -i\chi^a{}_{i}} \psi^a{}_{i\,0},\quad 
\bar\psi^a{}_{i} = \bar\psi^a{}_{i\,0} e^{-i\g_5 \th^a{}_{i} +i\chi^a{}_{i}}.
\eeq
By this local phase transformation, we are able to transcribe the bulk interaction into the 
boundary interaction. 
The local phase transformation and a finite scaling, bring us to the action which 
contains free field actions of the fermion and boson fields only as the bulk action
\beq
S &=& \frac{1}{2\pi} \int_M d\t d\s \sum_{a,i=1}^2\left[
\bar\psi^a{}_{i} \gamma \cdot \p \psi^a{}_{i} + \half\p\th^a{}_{i}\p\th^a{}_{i}\right] \nn\\
&&
+\frac{V_0}{4\pi}\int d\s \sum_{i=1}^2 \bar\psi^1{}_{i} e^{-\sqrt{2} i\g_5 \left(
\k_1 \th^1{}_i + \k_2 \th^2{}_i \right)} \psi^1{}_{i} ,
\eeq
where 
\beq
\k_1 = \sqrt{\frac{\a_1 -1}{2\a_1}}, ~~~ \k_2 = \sqrt{\frac{\a_2 -1}{2\a_2}}. 
\eeq
Since the boson action of the model of the hetero-junction is identical to the 
spin-dependent TL model with a single barrier except that the model of the hetero-junction has 
only one channel, the perturbation analyses of two models are similar to each other. 
For this reason, we may refer to ref.\cite{LeeU(1)2015} for details of the procedure.

Near the critical line, the boundary action may be expanded in $\k_1$ and $\k_2$ 
\beq \label{expansion}
S_{\rm boundary} &=& \frac{V_0}{4\pi}\int d\s \sum_{i=1}^2 \Bigl[\bar\psi^1{}_{i}\psi^1{}_{i}
-\sqrt{2} i \bar\psi^1{}_{i} \g_5 (\k_1 \th^1{}_i+ \k_2 \th^2{}_i) \psi^1{}_{i} \nn\\
&&~~~~~~~~~~~~~~~~~~~~~~~~~~ - \bar\psi^1{}_{i} \left( \k_1 \th^1{}_i+ \k_2 \th^2{}_i
\right)^2 \psi^1{}_{i} + \cdots \Bigr] \Bigl\vert_{\t=0} .
\eeq
The radiative corrections to the boundary hopping interaction can be calculated in the 
framework of the usual perturbation theory of a renormalizable field theory \cite{brown1992}.
Examining the fully interacting propagator of the fermion fields $\psi^1{}_i$, $i=1,2$ on the boundary, 
we can calculate the corrections to the boundary hopping interaction.
The fully interacting propagator of fermion field $\psi^1{}_i$ on the boundary is
defined as the bulk Green's function evaluated on the boundary
\beq \label{limit}
\bolG_{(i|\a\b)}(\s_1-\s_2) &=& \lim_{\t_1, \t_2 \rightarrow 0} 
\langle 0 \vert T \psi^1_{i\a}(\t_1,\s_1) \barpsi^1_{i\b}(\t_2,\s_2) \vert 0\rangle \nn\\
&=& \int D[\barpsi, \psi] 
\psi^1{}_{i\a}(0,\s_1) \barpsi^1{}_{i\b}(0,\s_2) \exp \left(-S_{\rm bulk}- S_{\rm boundary}\right) .
\eeq
Since the bulk and boundary actions are diagonal in the flavor indices $i$, we may omit the flavor indices 
$i$ hereafter. 
We treat the boundary hopping term as a part of interaction, even though it is a fermion bilinear. 
The first term in the expansion of the boundary action $S_{\rm boundary}$, Eq.(\ref{expansion}) contributes to the corrections of the fermion propagator at tree level, $\bolG^{(0)}_{(\a\b)}$
\begin{subequations}
\beq
\bolG^{(0)}_{(\a\b)} (\s_1-\s_2)
&=& \sum_n \bolG^{(0)}_{\a\b} [n] e^{in(\s_1-\s_2)}, \\
\bolG_{\a\b} [n] &=&  \frac{V_0^2}{16\pi^2} G_{\a\b}[n],~~~~~n \in \bbZ+1/2
\eeq
\end{subequations}
where $ G_{\a\b}[n]$ is the Fourier component (momentum number space representation) of the free fermion 
propagator on the boundary.

The free fermion Green's function on the boundary has two different Fourier decompositions, depending on
the direction, along which the limit Eq.(\ref{limit}) is taken. Throughout the paper we choose the 
retarded Green's function, which is given as 
\beq
G(\s_1-\s_2) &=& \sum_{n \in \bbZ+1/2} \frac{1}{2\pi} \g^0  \begin{pmatrix}
\th(n) & 0 \\ 0 & \th(-n) \end{pmatrix} e^{in(\s_1-\s_2)} , \label{freepropagator}
\eeq
where $\th(n)$ is the unit step function (Heaviside step function),
\beq
\th(n) = \begin{cases}  1 & \quad \text{for}~~ n \ge 0 \\
0 & \quad \text{for}~~ n < 0 ~~.
\end{cases} 
\eeq
Note that the fermion parpagator Eq.(\ref{freepropagator}) has only half integer modes in its Fourier expansion,
since the fermion fields satisfy the anti-periodic condition on a cyclidrical space. 

\begin{figure}[htbp]
   \begin {center}
    \epsfxsize=0.7\hsize
%
	\epsfbox{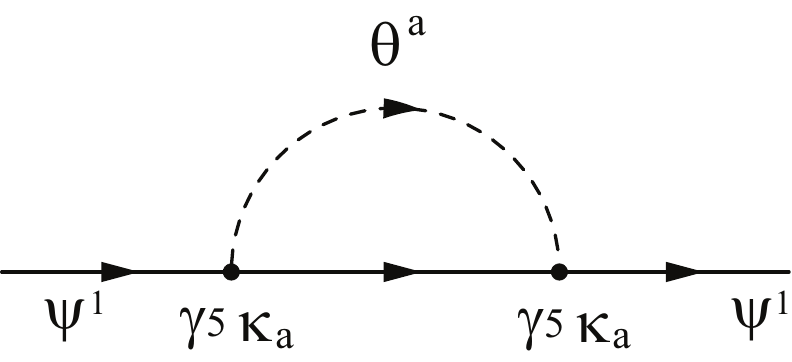}
   \end {center}
   \caption {\label{hopping} The first order correction to the fermion Green's function}
\end{figure}

Iterating the second term in the expansion of the boundary action $S_{\rm boundary}$, 
Eq.(\ref{expansion}) leads to the first-order one-loop correction to the Green's function. 
It corresponds to the Feynman diagram of Fig.\ref{hopping}. The result of the calculation is given as 
\beq
\bolG^{(1)}_{\a\b}[n] = \frac{V_0^2}{16\pi^2} \frac{1}{4} \left(2- \frac{1}{\a_1} - \frac{1}{\a_2}
\right)\left\{\frac{1}{\e} + \sum_{m=1/2}^\infty \frac{1}{m} + \text{finite terms} \right\} G_{\a\b}[n],
\eeq
where
$\bolG^{(1)}_{\a\b}[n]$ is the Fourier transformation of $\bolG^{(1)}_{\a\b}$
\beq
\bolG^{(1)}_{\a\b} [n] &=&  \int\frac{d\s}{2\pi} \bolG^{(1)}_{\a\b} (\s) e^{-in\s},
~~ ~~~~~n \in \bbZ+1/2 .
\eeq
The one-loop correction contain both infrared and ultraviolet divergences, which should be regularized. 
Since the one-loop correction is proportional to the tree level correction, we may absorb these divergences 
into the renormalization of the coupling constant.

\begin{figure}[htbp]
   \begin {center}
    \epsfxsize=0.5\hsize
%
	\epsfbox{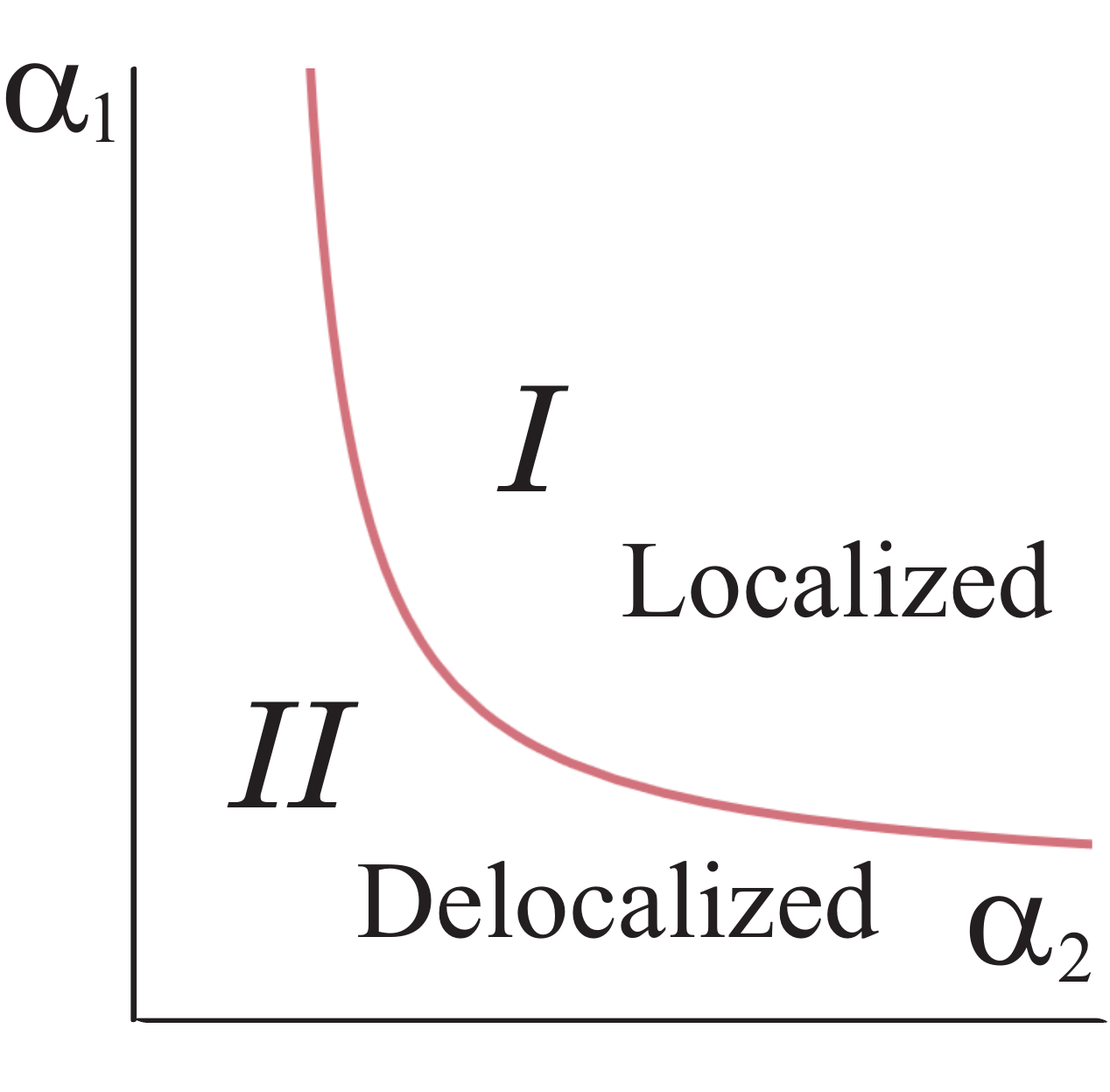}
   \end {center}
   \caption {\label{heterocr} The critical line and the phase diagram of the hetero-junction.}
\end{figure}

Up to the first order the radiative corrections to the Green's function is evaluated
in the momentum number space as 
\beq
\bolG^{(0)}_{\a\b} [n]+ \bolG^{(1)}_{\a\b} [n] = \frac{V^2_0}{16\pi^2}  \left\{1 + \frac{1}{4}
\left(2- \frac{1}{\a_1} - \frac{1}{\a_2}\right)\left(\frac{1}{\e}+ 
\zeta(1)\right)+\cdots \right\} G_{\a\b}[n],
\eeq
where $\zeta(1)$ is the Riemann zeta function. 
Renormalizing the coupling constant $V$, we may absorb the divergences
\beq
V^2 &=& V^2_0 \Biggl\{1+ \left(1- \frac{1}{2\a_1} - \frac{1}{2\a_2}\right) \ln \frac{\Lambda^2}{\m^2} \Biggr\} \nn\\
&=& V^2_0 \left(\frac{\Lambda^2}{\m^2} \right)^{1- \frac{1}{2\a_1} - \frac{1}{2\a_2}}.
\eeq
Where $1- \frac{1}{2\a_1} - \frac{1}{2\a_2} >0$ (the region I in Fig.\ref{heterocr}) 
the hopping interaction becomes a relevant operator and  
it tends to be strong in the zero temperature limit. 
In the region (the region II in Fig.\ref{heterocr}) where $1- \frac{1}{2\a_1} - \frac{1}{2\a_2} <0$  the 
hopping interaction is an irrelevant operator and it becomes weak in the zero temperature limit. 
We confirm that the point $(\a_1,\a_2) = (1,1)$ is on the critical line. 
If the two wires have the same TL parameter $\a_1=\a_2=\a_{\rm eff}$ (homo-junction), we would obtain the 
RG exponent as $1- \frac{1}{\a_{\rm eff}}$. Thus, the critical behavior of the hetero-junction 
with two different TL parameters $\a_1$ and $\a_2$ may be equivalent to that of the homo-junction with
a TL parameter $\a_{\rm eff}$ given as
\beq
\a_{\rm eff} = \frac{2}{\frac{1}{\a_1}+ \frac{1}{\a_2}}.
\eeq
This result is consistent with the previous works \cite{Safi95,Hou2012}. 

In the zero temperature limit, 
the boundary state in the weak coupling regime flows to the state $\vert N, N; D, D \rangle$
\begin{subequations}
\beq
\left(\phi^a_L - \phi^a_R \right) \vert N, N; D, D \rangle &=&0, ~~
\left(\bar\phi^a_L +\bar \phi^a_R \right) \vert N, N; D, D \rangle=0, ~~ a = 1, 2
\eeq
and in the strong coupling regime the boundary state flows to the state $\vert D, D; D, D \rangle$
\beq
\left(\phi^a_L + \phi^a_R \right) \vert D, D ; D, D \rangle &=& 0, ~~
\left(\bar\phi^a_L +\bar \phi^a_R \right) \vert D, D; D, D \rangle=0, ~~ a = 1, 2 .
\eeq
\end{subequations}
On the critical line the boundary state depends only on the coupling constant of the hopping interation $V$,
\beq
\vert B_{\rm critical} \rangle 
= \exp \left\{\frac{V}{4\pi} \int d\s \sum_{i=1}^2 \bar\psi^1{}_i \psi^1{}_i
\right\}\vert N, N; D, D \rangle. 
\eeq

\section{Particle-Kink Duality}

In the strong coupling regime, the boundary state becomes the Dirichlet state, the particles are mostly localized at the minima of the boundary periodic potential, which represents the hopping interaction between two wires. 
Electron may no longer transfer between two wires in the localized phase of the 
strong coupling regime. 
In the strong coupling regime of the periodic potential, the extended object called kink, which is a classical solution interpolating between local minima of the periodic potential, may become 
dynamical. Their collective motion may be described by the dual action, which takes exactly 
the same form as that of the particle fields but with weak coupling constants. 
It may requires a boundary kinetic term of the boson field for the kink to be a dynamical object. 
Such a kinetic term is absent at the tree level bare action. It is usually ignored in the 
perturbation theory, since the boundary kinetic term is an irrelevant operator.

\begin{figure}[htbp]
   \begin {center}
    \epsfxsize=0.3\hsize
%
	\epsfbox{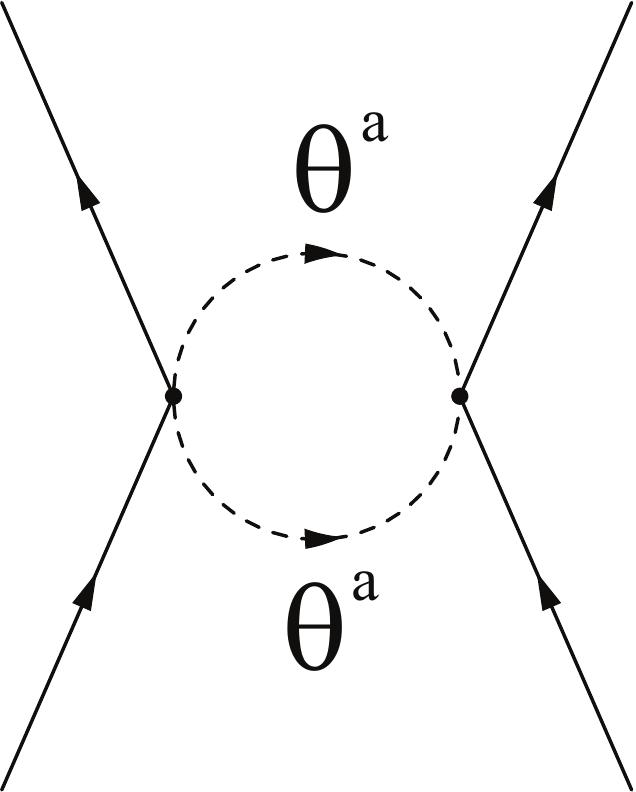}
   \end {center}
   \caption {\label{fourfermi} One-loop correction to the four-fermi interaction}
\end{figure}

However, the boundary kinetic term may be generated by the radiative corrections, which may not be ignored 
in the low energy regime. 
Since the boundary kinetic term of the boson theory is written as a four fermi term
\beq
\frac{M}{2} \left(\frac{d \phi}{d \s}\right)^2 := \frac{M}{2^3} \left(\barpsi_{11} \g^0 \psi_{11} - \barpsi_{12}\g^0 \psi_{12} \right)^2 .
\eeq
The one-loop radiative correction to the four-fermi Green's function on the boundary, which 
is depicted by Feynman diagram Fig.\ref{fourfermi}, may yield a finite boundary kinetic term 
\beq
\bolG_{(ijkl \vert \a \b\g \d)}(\s_1, \s_2, \s_3, \s_4) &=& \Bigl\langle\psi^1{}_{i\a}(0,\s_1) \barpsi^1{}_{j\b}(0,\s_2)  \psi^1{}_{k\g}(0,\s_3) \barpsi^1{}_{l\d}(0,\s_4) 
\Bigr\rangle .
\eeq

Taking into account the radiative corrections, we may write the boundary action for the boson field as 
\beq
S_{\rm boundary} = \int d\s \left\{ \frac{M}{2} \left(\frac{d\phi^1}{d\s}\right)^2 + 
\frac{V}{2} \left(e^{i\phi^1}+ e^{-i\phi^1} \right) \right\}. 
\eeq
The classical equation of motion is read as 
\beq
M \frac{d^2 \phi^1}{d\s^2} - V \sin \phi^1 = 0 ,
\eeq
and it is satisfied by the well-known kink solution
\beq
\phi^1(\s) =\phi_K(\s) = 2 \arccos \left[-\tanh\left(\sqrt{\frac{|V|}{M}}\s\right)\right] .
\eeq
This solution interpolates between two minima of the periodic potential at $\phi^1 =0,~ 2\pi$. 
Assuming that kinks are located far apart, we may write the multi-kink solution as 
\beq
\phi^1(\s) = \sum^{n}_{i=1} e_i\,  \phi_K(\s-\s_i), ~~~ e_i = 1,~ \text{or}~ -1 .
\eeq
The solution with $e =1$ corresponds to a kink and that with $e=-1$ to an anti-kink. 

The partition function in the kink sector may be written as 
\beq
Z &=& \int D[\phi^1,\phi^2] e^{-S[\phi^1,\phi^2]} \nn\\
&=& \sum_n \sum_{\{e_i\}} \frac{1}{n!} \int \prod_{i=1}^n d\s_i \, Z[n;\{e_i\};\s_1, \dots, \s_n], 
\eeq
where the partition function of the $n$-kinks sector is given by 
\begin{subequations}
\beq
Z[n;\{e_i\};\s_1, \dots, \s_n]&=& \int D[\phi^1,\phi^2] 
\,e^{-S_{\rm bulk}[\phi^1, \phi^2]-S_{\rm boundary}[\phi^1]}, \label{nsector}\\
\phi^1(\t=0, \s) &=& \sum_{i=1}^n e_i \phi_K(\s-\s_i) . \label{const}
\eeq
\end{subequations}
Introducing a Lagrangian multiplier $\l$ to impose the boundary condition Eq.(\ref{const}), 
we may rewrite the partition function of the $n$-kinks sector, abbreviated by
$Z[n]$, as follows
\beq
Z[n] &=& y^n_0 \int D[\phi^1,\phi^2] \int D[\l] \exp[-S_{\rm bulk} - iS_C ] , \nn\\
S_C &=& \int d\s \l(\s) \left(\phi^1(0,\s) - \phi_K(\s)\right).
\eeq
The boundary action evaluated with the $n$-kinks solution Eq.(\ref{const})
yields the instanton fugacity
\beq
S_{\rm boundary}[\phi^1] = -8 n \sqrt{M|V|}, ~~~y_0 = \exp\left[-8 \sqrt{M|V|} \right].
\eeq

In order to evaluate the partition function $Z[n]$ explicitly, we may rewrite it in terms of the 
Fourier modes of the boson fields 
\begin{subequations}
\beq
\phi^a(\t,\s) &=& \frac{1}{2\pi} \sum_n \int \frac{dq}{2\pi}\, \phi^a_n[q] \,e^{iq\t + in\s}, \\
\phi_K(\s) &=& \frac{1}{2\pi} \sum_n \phi_{K n} \, e^{-in\s} , \\
\l(\s) &=& \frac{1}{2\pi} \sum_n \l_n e^{in\s} .
\eeq
\end{subequations}
In terms of the Fourier modes, the bulk action and the constraint are read as 
\begin{subequations}
\beq
S_{\rm bulk} &=& \frac{1}{8\pi^2} \sum_n \int \frac{dq}{2\pi} \,\sum_{a,b =1}^2 \phi^a_n[q]
\left(q^2+n^2\right)\left(\d_{ab}+ g_{ab} \right) \phi^b_{-n} [-q], \\
S_C &=& \frac{1}{2\pi} \sum_n \l_n \left\{ \int \frac{dq}{2\pi} \phi^1_n[q] - \phi_{Kn} \right\} .
\eeq
\end{subequations}
Integrating out $\phi^a_n[q]$ and $\l_n$, we have 
\beq
Z[n] &=& y^n_0 \int D[\l] \exp \Biggl\{-\frac{1}{2} \sum_n \int \frac{dq}{2\pi}
\frac{1}{2} \left(\frac{1}{\a_1}+ \frac{1}{\a_2} \right) \l_n \frac{1}{q^2+n^2} \l_{-n} \nn\\
&&  ~~~~~+ \frac{i}{2\pi}\sum_n \l_n \phi_{Kn} \Biggr\} \nn\\
&=& y^n_0 \exp \left\{ - \frac{1}{(2\pi)^2} \sum_n |n| \left(\frac{2\a_1 \a_2}{\a_1+\a_2}\right) 
\phi_{K\,n} \phi_{K \,-n} \right\}
\eeq

In the strong coupling regime, $V/M \gg 1$, we can approximate $\phi_{Kn}$ as 
\beq
\phi_{Kn} &=& 2\pi i \sum_{i=1}^n e_i \, \frac{1}{n} e^{in\s_i}.
\eeq
Then it follows that 
\beq
Z &=& \sum_{n} \sum_{\{e_i\}} \frac{ y^{n}_0}{n!}\int \prod_{i=1}^n d\s_i  \nn\\
&& \exp\Biggl\{
-\frac{1}{2} \sum_n \frac{1}{|n|} \left(\sum_{i=1}^n e_i e^{in\s_i}\right) \left(\frac{2\a_1 \a_2}{\a_1+\a_2}\right) 
\left(\sum_{i=1}^n e_i e^{-in\s_i}\right) \Biggr\}.
\eeq
If we introduce a dual field, $\widehat \phi$
\beq
\whatphi(\s) = \sum_n \frac{1}{2\pi}\whatphi_n e^{in\s},  
\eeq
we may cast the partition function into the partition function of TL liquid 
\beq
Z &=& \sum_{n} \sum_{\{e_i\}} \frac{ y^{n}_0}{n!}\int D[\whatphi] \int \prod_{i=1}^n d\s_i \nn\\
&& \exp \Biggl\{ -\frac{1}{4\pi} \sum_n \frac{|n|}{2\pi} \whatphi_n \frac{1}{2} \left(\frac{1}{\a_1}+ 
\frac{1}{\a_2} \right) \whatphi_{-n} + i \sum_{i=1}^n e_i \whatphi(\s_i) \Biggr\} \nn\\
&=& \int D[\whatphi] \, \exp \Biggl\{ - \frac{{\widehat \a}}{4\pi} \int d\t d\s \p \whatphi \p \whatphi - \frac{\widehat V}{2\pi} \int d\s \left(e^{i\whatphi}+ e^{-i\whatphi} \right)\Bigl\vert_{\t=0}
\Biggr\}, 
\eeq
where 
\beq
{\widehat \a} = \frac{1}{2} \left(\frac{1}{\a_1}+ 
\frac{1}{\a_2} \right).
\eeq
We may conclude that in the strong coupling regime, if stable kinks exist, their collective motion 
may be described by a single TL liquid with a TL parameter given as $\whatphi$
\beq
S_{\rm Dual|} &=& \frac{{\widehat \a}}{4\pi} \int d\t d\s \p \whatphi\, \p \whatphi + \frac{\widehat V}{2\pi} \int d\s \left(e^{i\whatphi}+ e^{-i\whatphi} \right)\Bigl\vert_{\t=0}.
\eeq
It is interesting to 
note that the TL parameter of the dual field is just the inverse of the $\a_{\rm eff}$:
${\widehat \a} = 1/\a_{\rm eff}$. 

\begin{figure}[htbp]
   \begin {center}
    \epsfxsize=0.5\hsize
%
	\epsfbox{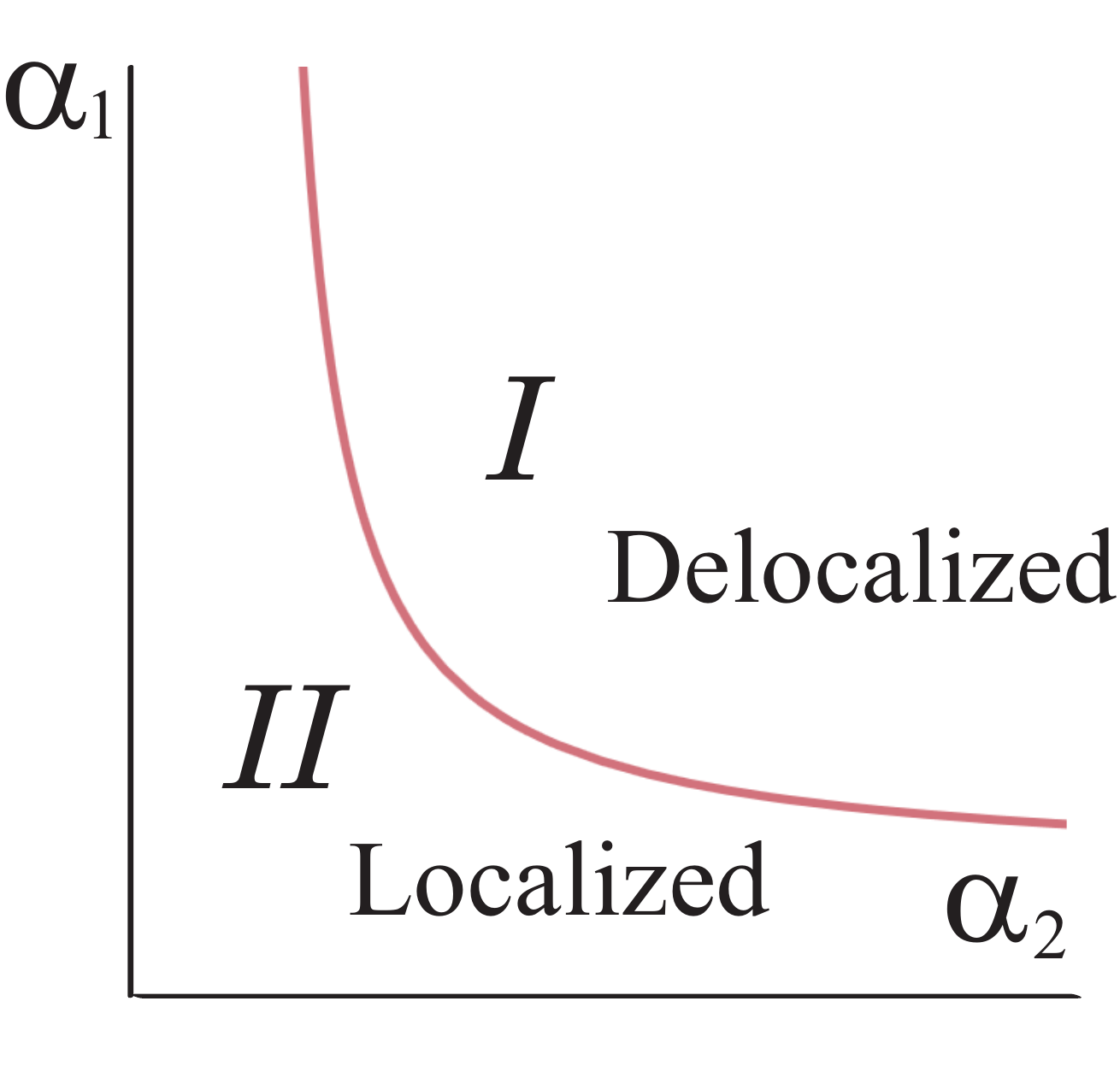}
   \end {center}
   \caption {\label{dualcr} The critical line and the phase diagram of the dual theory.}
\end{figure}

If we apply the perturbative analysis \cite{LeeU(1)2015}, 
similar to what we performed in the previous section, we 
would find the RG flow of the boundary interaction of the dual theory as 
\beq
{\widehat V}^2= {\widehat V}^2_0 
\left(\frac{\Lambda^2}{\mu^2}\right)^{\frac{{\widehat \a}-1}{2{\widehat \a}}} 
= {\widehat V}^2_0 \left(\frac{\Lambda^2}{\mu^2}\right)^{\frac{\a_1+\a_2-2\a_1 \a_2}{2(\a_1+\a_2)}} .
\eeq 
In the dual theory the hopping boundary interaction becomes an irrelevant operator in the region 
(region I in Fig.\ref{dualcr}) where
\beq
{\widehat \a} < 1, ~~~\text{equivalently}~~ \frac{1}{\a_1} + \frac{1}{\a_2} < 2 ,
\eeq
and a relevant operator in the region (region II in Fig.\ref{dualcr}) where 
\beq
{\widehat \a} >1, ~~~\text{equivalently}~~ \frac{1}{\a_1} + \frac{1}{\a_2} > 2.
\eeq
This is exactly opposite to the RG flow of the hopping interaction of the particle theory, depicted by
the phase diagram of Fig.\ref{heterocr} . Thus, the localized phase (strong coupling regime) of the 
particle theory is mapped completely onto the delocalized phase (weak coupling regime) of the dual theory
and vice versa. 
It is summarized by the phase diagram of the dual theory, Fig.\ref{dualcr}.

\section{Conclusions}

We studied the hetero-junction of two quantum wires, applying the boundary state formulation. 
It is the simplest building block of the quantum circuits, which deserves a detailed study. 
If the two quantum wires have different TL parameters, $\a_1$ and $\a_2$, the model becomes similar to the 
spin-dependent TL model with a single barrier. Still there are also some differences between 
two models. The model of hetero-junction has only one channel of electron transport while
the spin-dependent TL model has two channels. We performed a perturbative analysis to 
examine its critical behaviors. In the literatures the RG exponents of the similar models 
have been calculated by the Poor Man's Scaling method or its variants. But it is desirable 
to evaluate them by a more direct method. Applying the bosonization and refermionization to the 
model, we redefined the model as a Thirring model with a boundary fermion bilinear. 
Then we fransform away the interaction in the bulk action by axial and vector $U(1)$ local 
phase transformations. Since the boundary fermion bilinear breaks the $U(1)$ axial symmetry
explicitly, the corresponding boson degrees of freedom become dynamical. However, the interaction 
between the fermion fields and the local $U(1)$ phase boson fields take place only 
at the boundary and the bulk action contains only free field actions. It enables us to 
develop a perturbative expansion. The bulk free action defines the free propagators of the 
fermion and boson fields on the boundary and the radiative corrections to the 
boundary interaction can be calculated by usual Feynman diagram expansions. 
By evaluating a one-loop Feynman diagram, we obtained the RG exponent of the 
boundary hopping interaction: The critical line, which defines the phase boundary at 
zero temperature is determined by an effective TL parameter of the particle theory, 
$\a_{\rm eff} = \frac{2}{\a^{-1}_1+ \a^{-1}_2}$. This result confirms the previous one,
obtained by different analyses. 

We also examined the particle-kink duality of the model of hetero-junction. As in the 
case of the spin-dependent TL model with a single barrier, we expect that in the strong coupling 
regime, a new degree of freedom called kink, described by the dual field, may emerge. 
The boundary kinetic term of the boson field may be generated by radiative corrections to the 
Green's function of four fermion fields. Since the model of hetero-junction has only one channel
of transport, one kind of kink appears. As a result, the dual theory is described by a single 
TL liquid with a periodic boundary interaction. The TL parameter of the dual theory is found to be
the inverse of the effective TL parameter, $\a_{\rm dual} = 1/\a_{\rm eff}$. It follows from this 
relation that the strong coupling regime of the particle theory is completely mapped onto the 
weak coupling regime of the dual theory and vice versa: In the localized phase of the particle 
sector, the dual object becomes dynamical while in the delocalized phase of the particle sector
the dual object is localized. At this point the model of hetero-junction differs from the 
spin-dependent TL model. In the spin-dependent TL model, the localized phase of the particle 
theory does not exactly match that of the delocalized phase of the dual theory. 
The particle-kink duality, if exists, its consequence should be observed directly in 
experiments with junctions of the quantum wires. Extensions of the present work to more 
complex models will be discussed elsewhere.


\vskip 1cm

\noindent{\bf Acknowledgments}\\
This work was supported by Kangwon National University.


%

%





\end{document}